\documentclass{article}
\usepackage{spconf,amsmath,graphicx,hyperref}
\usepackage{amssymb}

\title{HRTF Upsampling Across Varying Measurement Configurations with Geometry-Aware Query-Conditioned Aggregation}
\name{
Xingyu Chen$^{1}$,
Hanwen Bi$^{1}$,
Sipei Zhao$^{1}$,
Fei Ma$^{1}$,
Eva Cheng$^{1}$,
Ian S. Burnett$^{2}$
}
\address{
$^{1}$Centre for Audio, Acoustics and Vibration,
University of Technology Sydney, Australia\\
$^{2}$Faculty of Information Technology,
Monash University, Australia
}

\begin{document}
\ninept
\maketitle

\begin{abstract}
Personalized head-related transfer functions (HRTFs) are essential for spatial audio rendering, but densely measuring an individual's HRTFs is costly and time-consuming.
HRTF upsampling reduces this burden by estimating dense HRTFs from sparse measurements.
Recent learning-based methods have achieved promising performance, but many remain tied to predefined measurement configurations.
In this work, we propose \textbf{GeoAtt}, a variable-context HRTF upsampling framework that uses a single trained model across varying measurement configurations.
GeoAtt performs geometry-aware, query-conditioned spatial aggregation over the available measurements independently at each frequency bin,
followed by frequency-domain modeling using Conformer blocks.
The relative geometry between the target and measured directions is incorporated as an additive bias in the cross-attention.
Experiments on the SONICOM dataset show that a single trained model achieves the lowest log-spectral distortion across all four canonical Listener Acoustic Personalization (LAP) challenge measurement configurations and further generalizes to configurations that are not explicitly included during training.
\end{abstract}


\section{Introduction}
Head-related transfer functions (HRTFs) characterize the direction-dependent acoustic filtering introduced by a listener's head, torso, and pinnae, and are therefore fundamental to spatial audio rendering~\cite{wenzel1993perceptual,keyrouz2007binaural,geronazzo2018we}.
Dense, personalized HRTFs are desirable, but their acquisition is costly and time-consuming~\cite{oberem2018experiments}. 
This motivates HRTF upsampling, which aims to estimate dense HRTFs from sparse measurements.

When only measurements from the target listener are available, HRTF upsampling is typically performed by exploiting spatial relationships among the measured directions.
Distance-weighted interpolation estimates HRTFs at unmeasured directions from
neighboring measurements~\cite{begault19943d,freeland2004interpositional},
while basis-function methods, particularly spherical harmonics (SHs), recover
the directional variation through spatial expansion
~\cite{zotkin2009regularized,ahrens2012hrtf,porschmann2019directional}.
Their performance, therefore, depends directly on the spatial information provided by the available measurements.

Learning-based methods additionally exploit regularities across listeners from
multi-subject HRTF datasets, providing a population-level prior for sparse HRTF
upsampling.
Representative methods include direction-conditioned autoencoders~\cite{ito2022head,ito2025spatial}, coordinate-based neural
fields~\cite{hrtf_field,lee2023global,masuyama2024niirf}, and a variety of
convolutional-, generative-, and Transformer-based architectures~\cite{zhao2025head,jiang2023modeling,hogg2024hrtf,hu2026hrtfformer}.
However, many learned models remain tied to predefined measurement
configurations, with fixed numbers or locations of input measurements embedded
in the learned spatial mapping.

\begin{figure}[t]
    \centering
    \includegraphics[width=0.5\textwidth]{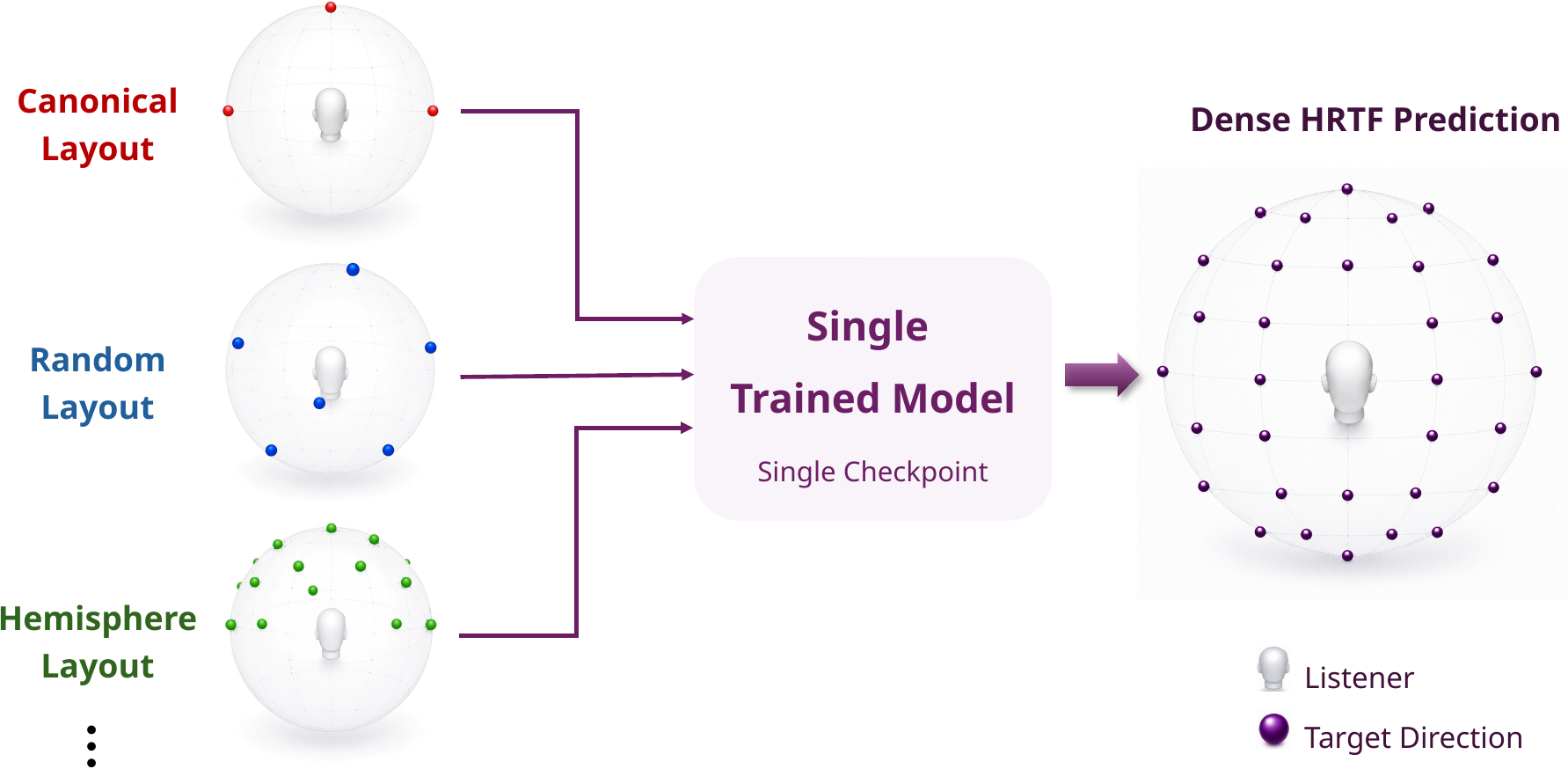}
    \caption{Illustration of variable-context HRTF upsampling across varying
measurement configurations.}
    \label{fig:task}
\end{figure}

Recent work has begun to relax this restriction.
SConvCNP~\cite{thuillier2024hrtf} predicts HRTFs from a variable-size set of
context measurements, demonstrating the feasibility of upsampling beyond
predefined measurement configurations.
Such flexibility is important in practical acquisition, where the number and locations of available measurements may vary because of acquisition
constraints, missing measurements, or adaptive sampling strategies.
We refer to this setting, where the available measurements may vary in both
number and spatial configuration, as variable-context HRTF upsampling illustrated in Fig.~\ref{fig:task}.

\begin{figure*}[t]
    \centering
    \includegraphics[width=\textwidth]{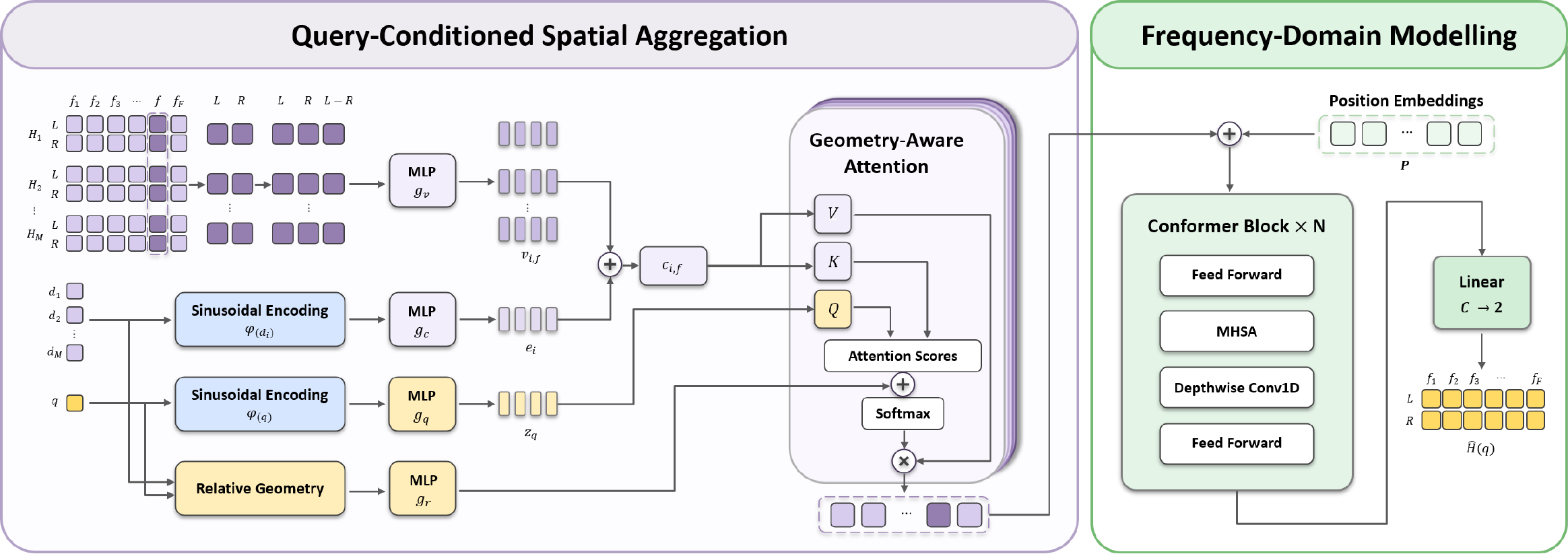}
    \caption{Overview of the proposed variable-context HRTF upsampling architecture.}
    \label{fig:architecture}
\end{figure*}

To address HRTF upsampling across varying measurement configurations, 
we propose \textbf{GeoAtt} to explicitly separate spatial aggregation from frequency-domain modeling.
For each target direction, geometry-aware cross-attention aggregates the
available measurements independently at each frequency.
The resulting query-conditioned frequency features are then processed by a
Conformer encoder, extending the frequency-domain modeling perspective of our
previous work~\cite{chen2026exploring} to the variable-context setting.
The main contributions are threefold:
First, we introduce a spatial--spectral framework that enables a single
model to operate on unordered measurements across varying measurement
configurations.
Second, we design a geometry-aware per-frequency spatial aggregation scheme
that uses the target direction as the cross-attention query and incorporates its relative geometry with the measured directions as an additive attention bias.
Third, we systematically evaluate a single trained model across diverse
measurement configurations, including configurations not explicitly included during training, demonstrating its flexibility across different measurement counts and spatial distributions.
Code and model weights will be released upon acceptance.

\section{Problem Formulation}
We consider far-field HRTFs, where source distance can be omitted, and denote $\mathbf {d}_i \in \mathbb {S} ^2$ as the $i$-th sound-source direction on the unit sphere.
We focus on HRTF magnitude upsampling, as the minimum-phase component can be
reconstructed from the magnitude response~\cite{kistler1992model}, while interaural time delay can be estimated separately with upsampling errors below the corresponding just-noticeable differences~\cite{zhao2025head}.
We therefore represent each binaural HRTF by its log-magnitude spectrum $\mathbf H(\mathbf d_i)\in\mathbb R^{2\times F}$,
where the two channels correspond to the left ($L$) and right ($R$) ears,
and $F$ denotes the number of frequency bins.

For each listener, the available measurements form a context set
\begin{equation}
\mathcal C=
\{(\mathbf d_i,\mathbf H_i)\}_{i=1}^{M},
\qquad
\mathbf H_i=\mathbf H(\mathbf d_i),
\label{eq:context}
\end{equation}
where both the number $M$ and the measurement directions $\{\mathbf d_i\}_{i=1}^{M}$ may vary across measurement configurations, as shown in Fig.~\ref{fig:task}.
Given $\mathcal C$ and a target direction $\mathbf q\in\mathbb S^2$, the objective is to predict the corresponding binaural log-magnitude HRTF
$\widehat{\mathbf H}(\mathbf q)\in\mathbb R^{2\times F}$.

\section{Method}

As illustrated in Fig.~\ref{fig:architecture}, GeoAtt factorizes
HRTF upsampling into query-conditioned spatial aggregation and frequency-domain modeling.
Given the context set
$\mathcal C=\{(\mathbf d_i,\mathbf H_i)\}_{i=1}^{M}$
and a target direction $\mathbf q$, we use $\mathbf q$ as the query in
cross-attention, while the available measurements provide the keys and values.
Spatial aggregation is performed independently at each frequency bin, producing
a query-conditioned feature for each frequency.
The resulting sequence of features is then modeled along the frequency axis
using Conformer blocks to predict the binaural HRTF at $\mathbf q$.

\subsection{Query-Conditioned Spatial Aggregation}

For each measurement $(\mathbf d_i,\mathbf H_i)\in\mathcal C$, the direction
$\mathbf d_i$ is encoded using a multiscale sinusoidal encoding~\cite{mildenhall2021nerf}, i.e.,
\begin{equation}
\phi(\mathbf d_i)
=
\left[
\mathbf d_i,\,
\{\sin(\pi 2^\beta\mathbf d_i),\,
\cos(\pi 2^\beta\mathbf d_i)\}_{\beta=0}^{B-1}
\right].
\label{eq:fourier}
\end{equation}
For the corresponding binaural HRTF $\mathbf H_i$, we construct at each
frequency bin $f$ the spectral feature
\begin{equation}
\mathbf s_{i,f}
=
[H_{L,i,f},\,H_{R,i,f},\,H_{L,i,f}-H_{R,i,f}]
\in\mathbb{R}^{3}.
\label{eq:binaural}
\end{equation}
The directional and spectral features are independently projected to the same
feature dimension and combined as
\begin{equation}
\mathbf c_{i,f}
=
g_v(\mathbf s_{i,f})
+
g_c(\phi(\mathbf d_i)),
\label{eq:contextfeature}
\end{equation}
where $g_v$ and $g_c$ are multilayer perceptron (MLP) shared across measurements.
The resulting context feature $\mathbf c_{i,f}$ is subsequently projected to the keys and values used for cross-attention at bin $f$.

The target direction $\mathbf q$ is first encoded by $\phi(\cdot)$ and then
projected to the same feature dimension by an MLP $g_q$ to obtain the query
representation
\begin{equation}
\mathbf z_q
=
g_q(\phi(\mathbf q)).
\label{eq:queryfeature}
\end{equation}

Rather than relying solely on standard cross-attention, we adopt the concept of relative positional encoding from Point Transformer~\cite{zhao2021point} to augment the attention matrix with relative geometric information.
Specifically, we define
\begin{equation}
\mathbf r_{q,i}
=
\left[
\mathbf q-\mathbf d_i,\;
\mathbf q^\top\mathbf d_i,\;
\arccos(\mathbf q^\top\mathbf d_i)
\right],
\label{eq:relative}
\end{equation}
where the three terms represent head-centered relative Cartesian displacement, angular similarity, and geodesic distance, respectively.
An MLP $g_r$ maps $\mathbf r_{q,i}$ to an additive attention bias.
The geometry-aware attention weight for measurement $i$ at frequency $f$ is
then
\begin{equation}
\alpha_{q,i,f}
=
\operatorname{softmax}_{i}
\left(
\frac{
(W_Q\mathbf z_q)^\top
(W_K\mathbf c_{i,f})
}{
\sqrt{d_a}
}
+
g_r(\mathbf r_{q,i})
\right),
\label{eq:spatial_attention}
\end{equation}
where $W_Q$ and $W_K$ are the query and key projections, respectively, and
$d_a$ denotes the attention dimension.
The context measurements are aggregated as
\begin{equation}
\mathbf t_{q,f}
=
\sum_{i=1}^{M}
\alpha_{q,i,f}
W_V\mathbf c_{i,f}
\in\mathbb R^{C},
\label{eq:spatial_aggregation}
\end{equation}
where $W_V$ denotes the value projection.

In practice, the geometry-aware cross-attention is extended to multi-head attention (MHA) following~\cite{vaswani2017attention}.
The aggregation is performed independently at each frequency bin, preserving the frequency axis for subsequent frequency-domain modeling.

\subsection{Frequency-Domain Modeling}

For each query direction $\mathbf q$, the aggregated features
$\{\mathbf t_{q,f}\}_{f=1}^{F}$ are stacked along the frequency axis as
\begin{equation}
\mathbf T_q
=
[\mathbf t_{q,1},\ldots,\mathbf t_{q,F}]^\top
\in\mathbb R^{F\times C}.
\label{eq:freq_sequence}
\end{equation}
A learnable frequency positional encoding
$\mathbf P\in\mathbb R^{F\times C}$ is added to distinguish the ordered
frequency bins, following the learnable absolute position
embeddings~\cite{devlin2019bert},
\begin{equation}
\mathbf X_q
=
\mathbf T_q+\mathbf P
\in\mathbb R^{F\times C}.
\label{eq:freq_pos}
\end{equation}

Building on the frequency-domain modeling principle of our previous
work~\cite{chen2026exploring}, $\mathbf X_q$ is processed by stacked Conformer
blocks~\cite{gulati2020conformer} along the frequency axis.
Unlike our previous work, which assumes a fixed measurement configuration, $\mathbf X_q$ is formed by query-conditioned aggregation of the available measurements.

Within each Conformer block, multi-head self-attention models long-range
dependencies across frequency bins, while a depth-wise one-dimensional convolution captures local spectral structure.
Together with feed-forward modules and residual connections, these operations
are repeated over $N$ blocks.
Finally, a shared linear projection maps the output to the left- and right-ear
log-magnitudes, yielding the predicted binaural HRTF $\widehat{\mathbf H}(\mathbf q)\in\mathbb R^{2\times F}$.

\subsection{Training Objective}

The model is trained in a supervised manner using log-spectral distortion (LSD)
and spectral gradient loss (SGL)~\cite{chen2026exploring}.
For $Q$ target directions, the LSD loss is
\begin{equation}
\mathcal L_{\mathrm{LSD}}
=
\frac{1}{2Q}
\sum_{j=1}^{Q}
\sum_{e=1}^{2}
\sqrt{
\frac{1}{F}
\sum_{f=1}^{F}
\left(
\widehat H_{j,e,f}-H_{j,e,f}
\right)^2
},
\label{eq:lsd_loss}
\end{equation}
where $\widehat H_{j,e,f}$ and $H_{j,e,f}$ denote the predicted and ground-truth
log-magnitudes, respectively.
The SGL constrains local spectral variation by penalizing the discrepancy
between adjacent-frequency differences,
\begin{equation}
\mathcal L_{\mathrm{SGL}}
=
\frac{1}{2Q(F-1)}
\sum_{j=1}^{Q}
\sum_{e=1}^{2}
\sum_{f=1}^{F-1}
\left|
\Delta_f\widehat H_{j,e,f}
-
\Delta_f H_{j,e,f}
\right|,
\label{eq:sgl_loss}
\end{equation}
where $\Delta_f$ denotes the first-order difference along frequency, e.g.,
$\Delta_f H_{j,e,f}=H_{j,e,f+1}-H_{j,e,f}$.
The total training objective is
\begin{equation}
\mathcal L
=
\mathcal L_{\mathrm{LSD}}
+
\mathcal L_{\mathrm{SGL}}.
\label{eq:training_loss}
\end{equation}

\section{Experiments}

\subsection{Dataset and Experimental Setup}

We evaluate GeoAtt on the SONICOM
dataset~\cite{engel2023sonicom}, which contains HRTFs from $200$ listeners
measured at $793$ directions.
The log-magnitude spectra between $187.5$~Hz and $19.875$~kHz are retained,
resulting in $F=106$ frequency bins for each ear.
The first $180$ listeners are used for training and model selection, while the
remaining $20$ are held out for testing.
We use a $160/20$ training--validation split to determine the model
configuration and training duration.
The final model is then retrained on all $180$ listeners for the selected
number of epochs and fixed for all subsequent evaluations.

During training, each sample is constructed by sampling a context set and a set of query directions.
With probability $0.5$, one of the four canonical measurement configurations defined in the Listener Acoustic Personalization (LAP)
challenge~\cite{hogg2025listener}, with $M\in\{3,5,19,100\}$, is selected
uniformly.
Otherwise, $M$ is sampled uniformly from $\{3,\ldots,100\}$, and $M$ measurement directions are selected randomly without replacement.
For each context set, $Q=64$ query directions are sampled from the remaining directions.
This strategy exposes the model to varying measurement configurations during training.

The model uses $4$ Conformer blocks with a feature dimension of $128$, a
feed-forward dimension of $256$, $8$ attention heads, and a convolution kernel
size of $7$.
The multiscale sinusoidal direction encoding uses $B=4$, and a dropout rate of
$0.1$ is applied throughout the network.
We use AdamW~\cite{loshchilov2018decoupled} with a learning rate of $1\times10^{-3}$, a weight decay of $1\times10^{-4}$, and a batch size of $8$.
The training duration is selected on the validation set, yielding $440$ epochs.
The final model is then trained from scratch for $440$ epochs on all $180$
training listeners using an NVIDIA L4 GPU.

At evaluation, all metrics are computed only over directions not included in the context set.
Performance is measured using LSD and interaural level difference (ILD) error~\cite{hogg2025listener}.

\begin{table*}[t]
\centering
\caption{Comparison on the canonical LAP measurement configurations.
LSD and ILD errors are reported in dB.
Configuration-specific methods use separate models, whereas variable-context methods use a single model across all configurations.}
\label{tab:canonical}

\setlength{\tabcolsep}{5pt}
\begin{tabular}{lcccccccc}
\hline
Method &
\multicolumn{2}{c}{$M=3$} &
\multicolumn{2}{c}{$M=5$} &
\multicolumn{2}{c}{$M=19$} &
\multicolumn{2}{c}{$M=100$} \\
& ILD & LSD & ILD & LSD & ILD & LSD & ILD & LSD \\
\hline
IOA3D~\cite{zhao2025head}
& $1.18\pm0.51$ & $4.67\pm0.42$
& $1.28\pm0.56$ & $4.61\pm0.54$
& $\mathbf{0.73}\pm0.25$ & $3.21\pm0.34$
& $0.56\pm0.10$ & $2.48\pm0.22$ \\

RANF~\cite{masuyama2025retrieval}
& $1.28\pm0.21$ & $4.53\pm0.39$
& $1.31\pm0.26$ & $4.47\pm0.52$
& $0.97\pm0.26$ & $3.65\pm0.42$
& $0.78\pm0.11$ & $3.03\pm0.35$ \\
\hline
SConvCNP~\cite{thuillier2024hrtf}
& $1.82\pm0.45$ & $4.90\pm0.35$
& $1.78\pm0.48$ & $5.47\pm0.52$
& $1.03\pm0.15$ & $3.71\pm0.22$
& $0.75\pm0.05$ & $2.58\pm0.19$ \\

GeoAtt
& $\mathbf{1.12}\pm0.32$ & $\mathbf{4.03}\pm0.36$
& $\mathbf{1.17}\pm0.48$ & $\mathbf{4.09}\pm0.46$
& $\mathbf{0.73}\pm0.19$ & $\mathbf{3.07}\pm0.30$
& $\mathbf{0.48}\pm0.10$ & $\mathbf{2.29}\pm0.22$ \\
\hline
\end{tabular}
\end{table*}

\begin{figure}[t]
    \centering
    \includegraphics[width=\columnwidth]{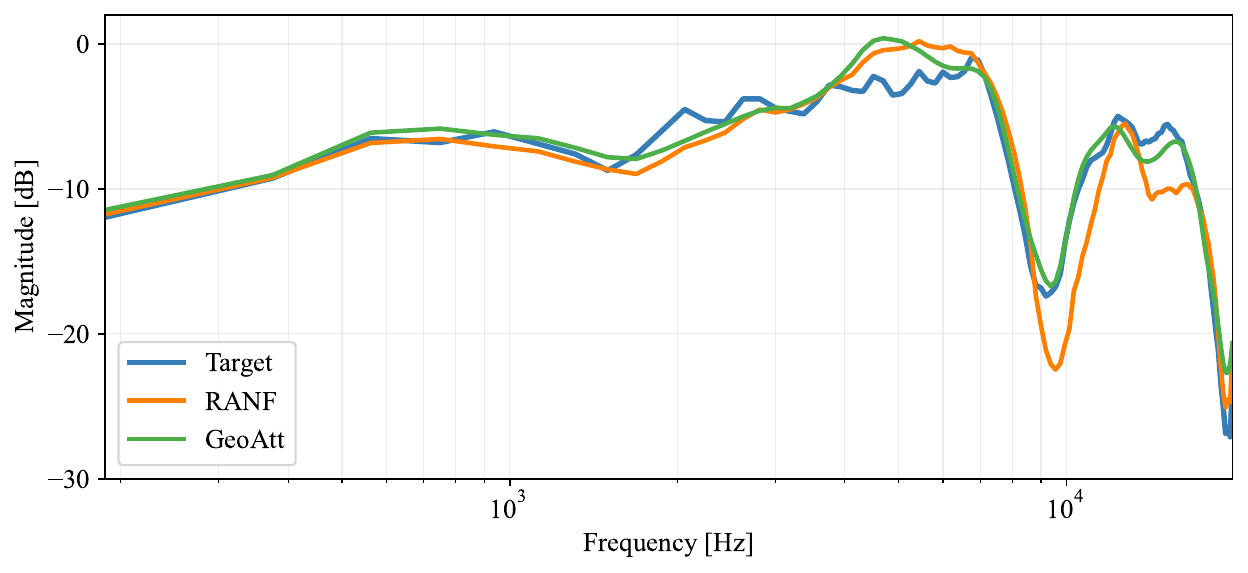}
    \caption{Example right-ear HRTF magnitude prediction for listener P0181 at the $90^\circ$ right-lateral direction with $M=19$ measurements.}
    \label{fig:spectrum}
\end{figure}

\begin{table}[t]
\centering
\caption{Generalization across varying measurement configurations at
$M=19$.}
\label{tab:config}

\setlength{\tabcolsep}{2.5pt}
\begin{tabular}{lcccc}
\hline
Configuration &
\multicolumn{2}{c}{SConvCNP} &
\multicolumn{2}{c}{GeoAtt} \\
& ILD & LSD & ILD & LSD \\
\hline
Lebedev-based
& $1.10\!\pm\!0.19$ & $3.85\!\pm\!0.24$
& $\mathbf{0.76}\!\pm\!0.20$ & $\mathbf{3.12}\!\pm\!0.28$ \\
$t$-design-based
& $1.09\!\pm\!0.18$ & $3.64\!\pm\!0.23$
& $\mathbf{0.74}\!\pm\!0.20$ & $\mathbf{3.05}\!\pm\!0.26$ \\
Hemisphere
& $1.25\!\pm\!0.16$ & $4.14\!\pm\!0.21$
& $\mathbf{0.78}\!\pm\!0.20$ & $\mathbf{3.31}\!\pm\!0.26$ \\
Horizontal ring
& $1.87\!\pm\!0.19$ & $5.26\!\pm\!0.33$
& $\mathbf{1.05}\!\pm\!0.24$ & $\mathbf{4.29}\!\pm\!0.31$ \\
\hline
\end{tabular}
\end{table}

\subsection{Comparison with Existing Methods}

We compare GeoAtt with existing HRTF upsampling methods under
two settings. 
First, we consider the canonical measurement configurations
defined in the LAP challenge~\cite{hogg2025listener}, with
$M\in\{3,5,19,100\}$. 
Second, we evaluate varying measurement configurations to assess generalization without configuration-specific retraining.

We compare against RANF~\cite{masuyama2025retrieval} and
IOA3D~\cite{zhao2025head}, the first- and second-ranked HRTF upsampling methods
in the LAP challenge, as configuration-specific baselines.
Both methods are retrained using their official implementations and
configurations. 
SConvCNP~\cite{thuillier2024hrtf} is included as a variable-context baseline and retrained using the same context-sampling
strategy as the proposed method.
All methods use the same SONICOM data split, HRTF preprocessing, frequency
range, and evaluation directions.

Table~\ref{tab:canonical} reports the results under the four canonical LAP
configurations. RANF and IOA3D use separately trained models for different
measurement configurations, whereas SConvCNP and GeoAtt use a
single trained model across all values of $M$.
The reported mean and standard deviation are computed across the 20 test
listeners.
Among the compared methods, GeoAtt achieves the lowest mean LSD
across all four configurations.

Compared with the best configuration-specific baseline, GeoAtt LSD by $0.50$, $0.38$, $0.14$, and $0.19$~dB for $M=3$, $5$, $19$,
and $100$, respectively, with the largest gains under sparse measurements.
It also achieves the lowest ILD error in three configurations and matches IOA3D in the remaining one.
Although SConvCNP handles varying configurations with a single trained model, it exhibits a performance gap compared with configuration-specific methods.
GeoAtt avoids this trade-off, reducing LSD over SConvCNP by up to $1.38$ dB, with the largest gains under sparse measurements.
Fig.~\ref{fig:spectrum} provides a representative example of the predicted HRTF magnitude. 
GeoAtt reproduces the target spectral structure more closely than RANF in this example, particularly around the high-frequency notches.
Overall, these results show that the proposed method combines variable-context flexibility with superior performance over the compared configuration-specific methods.

Table~\ref{tab:config} further compares the variable-context methods across four structured measurement configurations with $M=19$.
These configurations represent different established spherical sampling and HRTF measurement schemes: Lebedev-based and spherical $t$-design-based configurations provide well-distributed full-sphere coverage ~\cite{porschmann2019directional,hardin1996mclaren}, whereas upper-hemisphere
and horizontal-ring configurations provide progressively more restricted
spatial coverage~\cite{takane2002database,sridhar2017database}.
All configurations are instantiated using the available SONICOM measurement directions and are not explicitly included as structured configurations during training.
The same single trained model is evaluated across all four configurations.

The proposed method consistently outperforms SConvCNP in both LSD and ILD.
Both methods perform best under the well-distributed full-sphere configurations and degrade as spatial coverage becomes more restricted, with the horizontal ring being the most challenging condition.
Compared with SConvCNP, GeoAtt reduces LSD by $0.59$--$0.97$~dB, with the largest gains under the hemisphere and horizontal-ring configurations.
The ILD results exhibit the same trend, with larger improvements under the more spatially restricted configurations.
These results demonstrate generalization to structured measurement configurations not explicitly included during training, including substantial changes in spatial distribution and coverage.

\begin{table}[t]
\centering
\caption{Ablation study under the canonical LAP and hemispherical
configurations with $M=19$.}
\label{tab:ablation}

\begin{tabular*}{\columnwidth}{@{\extracolsep{\fill}}lcccc}
\hline
&
\multicolumn{2}{c}{Canonical LAP} &
\multicolumn{2}{c}{Hemisphere} \\
Variant & ILD & LSD & ILD & LSD \\
\hline
Mean pooling
&$1.18$ & $4.13$ & $1.19$ & $4.22$\\

w/o relative geometry
& $0.85$ & $3.24$ & $0.96$ & $3.48$ \\

w/o frequency modeling
& $1.01$ & $3.60$ & $1.04$ & $3.84$ \\

Conv-only
& $0.77$ & $3.21$ & $0.89$ & $3.48$ \\
\hline
Full model
& $\mathbf{0.73}$ & $\mathbf{3.07}$ & $\mathbf{0.78}$ & $\mathbf{3.31}$ \\
\hline
\end{tabular*}
\end{table}

\subsection{Ablation Study}

We conduct ablation experiments on four variants of GeoAtt:
(1) \textbf{Mean pooling} replaces query-conditioned cross-attention with
averaging over the context measurements;
(2) \textbf{w/o relative geometry} removes the geometric attention bias while
retaining cross-attention;
(3) \textbf{w/o frequency modeling} removes the Conformer and predicts each
frequency bin independently; and
(4) \textbf{Conv-only} retains one-dimensional convolution along the frequency
axis but removes global self-attention.
All variants follow the same training protocol and are evaluated under the
canonical LAP and hemispherical configurations with $M=19$.

Table~\ref{tab:ablation} shows that the full GeoAtt model achieves the best performance
under both configurations. 
Replacing query-conditioned cross-attention with mean pooling causes the largest degradation in spatial aggregation,
increasing LSD by $1.06$ and $0.91$~dB and ILD by $0.45$ and $0.41$~dB, respectively.
Removing the relative-geometry bias results in a smaller but consistent degradation, increasing LSD by $0.17$~dB in both cases. 
These results indicate that query conditioning provides the primary benefit for spatial aggregation, with relative geometry offering an additional, consistent improvement.
Removing frequency-domain modeling increases LSD by $0.53$~dB under both
configurations, whereas the Conv-only variant reduces this gap to $0.14$ and
$0.17$~dB, respectively.
This indicates that local spectral modeling accounts for a substantial portion
of the improvement, 
while the additional gain from the full Conformer supports the benefits of long-range spectral modeling.

\section{Conclusion}

This work addressed HRTF upsampling across varying measurement configurations
without requiring a separately trained model for each configuration.
Experiments on SONICOM show that GeoAtt reduces LSD by an average of $0.30$~dB over the strongest configuration-specific baselines across the four canonical LAP configurations, while also generalizing to measurement configurations not explicitly included during training.
The ablation study further shows that query-conditioned aggregation provides the primary benefit in spatial modeling, with additional gains from relative
geometry and explicit frequency-domain modeling.
Overall, the results support a flexible HRTF upsampling framework that can
accommodate changes in both the number and spatial distribution of available
measurements.

{\footnotesize
\bibliographystyle{IEEEbib}
\bibliography{refs}
}

\end{document}